\documentstyle[epsf]{l-aa}

\newcommand{\beq}{\begin{equation}}
\newcommand{\eeq}{\end{equation}}
\newcommand{\PhiPN}{{V_{_{\rm{PW}}}}}
\newcommand{\Deltat}{{\bar{t}}}
\newcommand{\Lambdaf}{{\Lambda_{\rm{f}}}}
\newcommand{\half}{{\textstyle\frac{1}{2}}}
\newcommand{\const}{{\rm const}}
\def\spose#1{\hbox to 0pt{#1\hss}}
\def\lta{\mathrel{\spose{\lower 3pt\hbox{$\mathchar"218$}}
     \raise 2.0pt\hbox{$\mathchar"13C$}}}
\def\gta{\mathrel{\spose{\lower 3pt\hbox{$\mathchar"218$}}
     \raise 2.0pt\hbox{$\mathchar"13E$}}}

\begin{document}

\thesaurus{02(02.01.2; 02.02.1)}

\title{Pseudo-Newtonian models of a rotating black hole field}
\author{Old\v{r}ich Semer\'ak\inst{1} \and
        Vladim\'{\i}r Karas\inst{2}}
\offprints{O.~Semer\'ak}

\institute{Department of Theoretical Physics,
           Faculty of Mathematics and Physics,
           Charles University, V~Hole\v{s}ovi\v{c}k\'ach~2,
           CZ-180\,00~Praha, Czech Republic;
           E-mail: semerak@mbox.troja.mff.cuni.cz
            \and
           Astronomical Institute,
           Faculty of Mathematics and Physics,
           Charles University, V~Hole\v{s}ovi\v{c}k\'ach~2,
           CZ-180\,00~Praha, Czech~Republic;
           E-mail: karas@mbox.troja.mff.cuni.cz}

\date{Received May ~~~~~, 1998; accepted}

\maketitle
\markboth{O.~Semer\'ak \& V.~Karas:
          Pseudo-Newtonian models of black holes}
         {O. Semer\'ak \& V. Karas:
          Pseudo-Newtonian models of black holes}

\begin{abstract}
  A pseudo-Newtonian description of the gravitational field which yields
  the equations of motion resembling, as closely as possible, the
  geodesic equation of general relativity is found for the Kerr
  spacetime. The potential obtained consists of three parts, interpreted
  as a purely Newtonian (gravitoelectric) term, a dragging
  (gravitomagnetic) term, and a space-geometry correction. The accuracy of
  the pseudo-Newtonian model is studied by a method which compares,
  systematically, two large sets of trajectories: geodesics in the Kerr
  spacetime versus test-particle trajectories in the pseudo-Newtonian
  field. It is suggested that every pseudo-Newtonian model should be
  submitted to analogous systematic analysis before it is used in
  astrophysical applications. A modified Newtonian potential which
  accounts for the frame-dragging effects can be a practical tool in
  studying stationary accretion discs. Non-stationary configurations are
  more complicated and we will not suggest to use this approach to such
  topics in accretion theory as gravitomagnetic oscillations of discs
  and their relation to quasi-periodic sources.

\end{abstract}

\section{Introduction}
\label{Intro}
Most of theoretical astronomy uses Newtonian theory of gravitation,
considering that the effects of general relativity are weak for
most astronomical objects. Even in situations where these effects do
become important or even dominant, pseudo-Newtonian models have
often been applied which aspire to mimic the corresponding relativistic
situations. Pseudo-Newtonian approaches aim to save numerical work, to
enable analytical solutions where fully relativistic treatment is too
cumbersome, and to provide new insights into the relativity theory and
its implications.

The present contribution concerns pseudo-Newtonian models developed in
order to describe very compact objects which presumably reside in the
nuclei of galaxies and in some X-ray binaries (e.g., Rees 1998). In
these sources, a key feature is an accretion disc around a rotating
black hole. The actual accretion flows are most likely non-stationary,
non-axisymmetric, and described by complex local physics. For the sake
of simplicity, however, a standard model of disc accretion (see Kato,
et al. (1998) for a textbook exposition of the accretion
theory including its recent advances) provides an ingenious
approximation: the standard disc is smooth, axially symmetric,
geometrically thin, and characterized by three parameters. It is
astrophysically realistic only in a restricted range of its parameters.
The pseudo-Newtonian approach has been devised in order to introduce
effects of general relativity into accretion models, especially
in the case of geometrically thick discs (Paczy\'nski \& Wiita 1980).
The black hole is treated there as a Newtonian body whose gravitational
field is determined by the potential $\PhiPN=-M/(r-2M)$ (geometrized
units with $c=G=1$ will be used throughout the paper). This simple
expression mimics the gravitational field of a {\em{}non-rotating\/}
(Schwarzschild) black hole quite accurately, in particular, it
reproduces correctly the radii of the marginally stable and the
marginally bound circular orbits of free test particles, $r_{_{\rm
ms}}=6M$ and $r_{_{\rm mb}}=4M$. Recently, Abramowicz et al.\ (1996)
proposed a certain rescaling of velocities in the field of $\PhiPN$ to
obtain better agreement with corresponding relativistic values. In this
way, calculations of the observed spectra of the discs could also be
improved. It has been argued, however, both on theoretical grounds
(Bardeen 1970) and from observations (Iwasawa et al.\ 1996; Karas \&
Kraus 1996), that the central black holes in galactic nuclei would
be more likely to rotate rapidly. Thus the main feature that should be included in
the pseudo-Newtonian models is the rotation of the central object. In
addition, the rotation-induced dragging must be taken into account when
studying non-equatorial warped discs around compact objects (Bardeen \&
Petterson 1975).

The pseudo-Newtonian potential for a non-rotating black hole has been
used in numerous works and we recall at least a few of them. For example,
Abramowicz et al. (1988) used $\PhiPN$ in their
introductory work on slim discs. Okazaki et al. (1987) adopted
this approach to describe global trapped oscillations of relativistic
accretion discs. Later, Nowak \& Wagoner (1991) devised another form of
the potential which is better suited for their purpose because it reproduces
the epicyclic frequency $\kappa$ with higher accuracy than $\PhiPN$.
Szuszkiewicz \& Miller (1998) studied the limit-cycle behaviour of
thermally unstable flows, also with the help of the pseudo-Newtonian
description. On quite a different subject, Daigne \& Mochkovitch (1997)
applied $\PhiPN$ to study the runaway instability in accretion discs
which could trigger gamma ray bursts. Very recently, Ruffert \& Janka
(1998) used $\PhiPN$ in a detailed study of neutron-star mergers and
related production of the bursts. It is quite apparent from this short
list of different astrophysical applications that the approach has its
advantages and enables qualitative (and quantitative, with some caution)
studies of different topics concerning black holes. Especially the
problems of accretion can be treated in this manner. It is understood
that {\it{}quantitative results\/} need always be checked carefully
within the full relativistic framework. This is particularly true for
time-dependent phenomena (like oscillations and waves in fluids) and for
computations of observed spectra. Notice that a textbook overview of the
whole subject can be found in Kato et al.\ (1998).

Now we come to the case of a rotating black hole. At least two topics of
current astrophysical interest can immediately be mentioned where the
frame-dragging effects around a rotating black hole are important: the
problem of oscillations of relativistic accretion discs (see Wagoner
1998 for a recent review), and that of precessing discs in low-mass
X-ray binaries (Stella \& Vietri 1997) and black-hole binaries (Wei et
al. 1998). It is apparent that the whole subject of oscillation modes in
relativistic discs calls for detailed investigation (cf.\ also
Markovi\'c \& Lamb 1998) and a suitable pseudo-Newtonian formulation can
be an appropriate tool before embarking on a complete solution. However,
unlike the case of stationary gaseous configurations, nonstationary
phenomena require a more refined choice of the pseudo-Newtonian model
because there are additional quantities apart from the location of
marginal circular orbits which must be correctly modelled
(e.g.\ $\kappa$).

Although the ever increasing computational facilities will make
practical reasons for the pseudo-Newtonian approach rather old-fashioned
in near future, what still remains desirable is a simple expression
which would simulate the rotating (Kerr) black hole accurately. This is
however difficult to find. For example, the potential suggested by
Chakrabarti \& Khanna (1992) could be useful in studying thin accretion
discs around rotating holes, but it applies only to the equatorial plane
and its interpretation is rather unclear (free parameters are fitted in
a purely pragmatic manner for a restricted set of trajectories). Another
form of the pseudo-Newtonian potential has recently been proposed by
Artemova et al. (1996). This pseudo-potential reproduces
very well steady circular motion but it ignores all effects which in the
true Kerr metric are ascribed to the Lense-Thirring precession and which
make test-particle trajectories non-planar. This is also the main reason
why, in contrast to the non-rotating case, pseudo-potentials for the
Kerr metric have had rather restricted impact. Also, a proper understanding
of these generalizations may be as difficult as employing general relativity
fully. However, the motivation which stems from attempts to
understand and interpret predictions of the relativity theory has not
disappeared.

It is the aim of our present contribution to discuss the
pseudo-Newtonian modelling of the gravitational field of a rotating
(Kerr) black hole, and also to propose how to test such models. One
should remember that the idea of the pseudo-Newtonian approach
represents a certain mathematical model rather than a rigorously defined
approximation such as e.g.\ the weak-field approximation of the Einstein
equations. The model does not aspire to represent any gravitational
theory and, in particular, the corresponding potential is not required
to satisfy any field equations. (It is exactly the lack of precise
definition of the approximation method which leads us to introduce a
test of accuracy of the model in the present article.) This fact,
however, does not diminish the practical value of Paczy\'nski-Wiita
potential and other pseudo-Newtonian models, which apply even to regions
with strong gravity and capture qualitative features of motion near the
horizon.

We start by writing down the spatial components of the Kerr geodesic
equation in Boyer-Lindquist coordinates $x^{\mu}=(t,r,\theta,\phi)$:
\begin{eqnarray}
  \Sigma\ddot{r}&=&
   M\Delta\Sigma^{-2}(\Sigma-2r^{2})
   (\dot{t}-a\dot{\phi}\sin^{2}\theta)^{2}
    \nonumber \\ &~&
   +[(r-M)\Sigma/\Delta-r]\,\dot{r}^{2}
   +a^{2}\dot{r}\dot{\theta}\sin2\theta  \nonumber \\ &~&
   +r\Delta(\dot{\theta}^{2}
   +\dot{\phi}^{2}\sin^{2}\theta)
        \, , \label{Kerr-r} \\
  \Sigma\ddot{\theta}&=&
   \{2Mr\Sigma^{-2}[a\dot{t}-(r^{2}
   +a^{2})\dot{\phi}]^{2}
   +a^{2}\dot{\theta}^{2}-
   \nonumber \\ &~&
   a^{2}\dot{r}^{2}/\Delta
   +\Delta\dot{\phi}^{2}\}\sin\theta\cos\theta
   -2r\dot{r}\dot{\theta} \, , \label{Kerr-theta} \\
  {\half}\Delta\Sigma^{2}\ddot{\phi}&=&
    Ma(\Sigma-2r^{2})(\dot{t}-a\dot{\phi}\sin^{2}\theta)\dot{r}
        \nonumber \\ &~&
   +2Mr\Delta[a\dot{t}
   -(r^{2}+a^{2})\dot{\phi}]\,\dot{\theta}
    \cot\theta               \nonumber \\ &~&
   -\Sigma(\Sigma-2Mr)(r\dot{r}+\Delta\dot{\theta}\cot\theta)
    \dot{\phi} \, , \label{Kerr-phi}
\end{eqnarray}
where
\begin{equation}
  \Delta=r^{2}-2Mr+a^{2},\quad\Sigma=r^{2}+a^{2}\cos^{2}\theta;
\end{equation}
$M$ and $a$ are parameters of the Kerr solution, and the dot denotes
differentiation with respect to the affine parameter normalized so that
the 4-momentum is $p^{\mu}=\dot{x}^{\mu}$.

In the next section, Newtonian equations of test-particle motion
in an axially symmetric gravitational field are given. In Sect.\
\ref{potentKerr}, we derive a pseudo-Newtonian potential for
which these equations get a form very similar to the above geodesic
equation in the Kerr spacetime.\footnote{For
a static and spherically symmetric field (both gravitational
and non-gravitational), a similar problem was solved by Jaen
(1967) using the Hamilton-Jacobi theory.}
In Sect.\ \ref{test}, numerical integration is carried out for a large
number of trajectories both in the Kerr and in the ``pseudo-Kerr''
fields. A specific criterion analogous to the method of Lyapunov
coefficients is introduced and the rate of divergence of each couple of
corresponding trajectories is determined numerically. The mean values
of this rate obtained for various combinations of constants of motion
indicate the quality (i.e.\ accuracy, as defined below) of the
pseudo-Newtonian approximation.

\section{Stationary and axisymmetric field}
\label{potentaxi}
In Euclidean space and in ellipsoidal coordinates
$x^{i}=(r,\theta,\phi)$, related to the Cartesian coordinates $(x,y,z)$
by
\begin{eqnarray}
  x&=&\sqrt{r^{2}+a^{2}}\,\sin\theta\cos\phi,  \nonumber \\
  y&=&\sqrt{r^{2}+a^{2}}\,\sin\theta\sin\phi,  \nonumber \\
  z&=&r\,\cos\theta,                           \nonumber
\end{eqnarray}
the kinetic energy of a particle of mass $m$ is
\begin{eqnarray}
 T &=& \frac{p^{2}}{2m}
    =\frac{\dot{x}^{2}+\dot{y}^{2}+\dot{z}^{2}}{2m} \nonumber \\
   &=&\frac{1}{2m}
    \left[\Sigma\left(\frac{\dot{r}^{2}}{r^{2}+a^{2}}
    +\dot{\theta}^{2}\right)
    +(r^{2}+a^{2})\,\dot{\phi}^{2}\sin^{2}\theta\right]\,,
\end{eqnarray}
where $a$ is a non-negative constant of the dimension of length and the
dot denotes differentiation with respect to the time parameter $t/m$
(normalized so that the momentum is
$\protect\vec{p}=\dot{\protect\vec{x}}$).
The motion of the particle in a stationary axisymmetric potential
$U=U(r,\theta,\dot{r},\dot{\theta},\dot{\phi})$ is described by
equations
\begin{eqnarray}
  \Sigma\ddot{r}&=&
    m^{2}(r^{2}+a^{2})
    [(U_{,\dot{r}})^{\displaystyle\cdot}-U_{,r}]
   \nonumber \\ &~&
   -r(a\dot{r}\sin\theta)^{2}/(r^{2}+a^{2})
   +a^{2}\dot{r}\dot{\theta}\sin2\theta
    \nonumber \\ &~&
   +r(r^{2}+a^{2})(\dot{\theta}^{2}+\dot{\phi}^{2}\sin^{2}\theta)
   \, , \label{axisym-r} \\
  \Sigma\ddot{\theta}&=&
    m^{2}[(U_{,\dot{\theta}})^{\displaystyle\cdot}-U_{,\theta}]
   +[a^{2}\dot{\theta}^{2}-a^{2}\dot{r}^{2}/(r^{2}+a^{2})
    \nonumber \\ &~&
   +(r^{2}+a^{2})\dot{\phi}^{2}]\,\sin\theta\cos\theta
   -2r\dot{r}\dot{\theta} \, , \\
  (r^{2}+a^{2})\ddot{\phi}&=&
   m(U_{,\dot{\phi}})^{\displaystyle\cdot}\sin^{-2}\theta
   \nonumber \\ &~&
  -2\,[r\dot{r}+(r^{2}+a^{2})\dot{\theta}\cot\theta]\,\dot{\phi}
   \, , \label{axisym-phi}
\end{eqnarray}
where we denote, for example,
\beq
\left(U_{,\dot{r}}\right)^\cdot=m\,\frac{\rm d}{{\rm d}t}
 \left(\frac{\partial U}{\partial\dot{r}}\right).
\eeq

\section{Potential for the Kerr field}
\label{potentKerr}
Relativistic gravitational fields can be interpreted as consisting of
three parts: the Newtonian (also Coulomb or gravitoelectric) component,
the dragging (gravitomagnetic) component, and the space-geometry
component. The Newtonian component is generated by mass-density and is
given essentially by the gradient of the $g^{00}$ component of the spacetime
metric. The dragging component is generated by mass-currents and
determined by curl of $g_{0i}$. The space-geometry component (determined
by $g_{ij}$) has no classical analog. This view and, in particular, the
analogy with electromagnetism are most straightforward within the
linearized approximation, but can be given a precise and invariant
meaning even in strong fields (Thorne et al. 1986; Jantzen
et al. 1992, and references cited therein). Let us compose the
potential for the Kerr spacetime in a way that acknowledges this
approach.

\subsection{The Newtonian component}
\label{potentKerr-Newt}
Cutting the Kerr manifold at $r=0$, one can obtain the spacetime which is
free of causality-violating regions. The cut leads to jumps
in derivatives of the metric at $r=0$ which are interpreted as
a thin massive layer. This induced mass spreads over the $r=0$
hypersurface and consists of the attractive singular ring ($z=0$,
$\rho_{_{\rm f}}=\sqrt{x^{2}+y^{2}}=a$) of infinite positive mass
density, spanned by the repulsive disc ($z=0$, $0\leq\rho_{_{\rm f}}<a$)
of negative mass density. Constructing the causally maximal extension of
the Kerr metric, Keres (1967) and Israel (1970) showed that the
Newtonian field generated by the massive layer corresponds to the
potential\footnote
{Cf.\ also Qadir (1986) where an electrically charged generalization
 of this potential was discussed. A somewhat different expression,
 $V=-(M/a)\,\arctan(a/r)$, was obtained by Krasi\'nski (1980) in
 construction of a Newtonian model of the Kerr source. For the
 other result, the potential of an oblate spheroidal homeoid
 $V=-(M/a){\rm arccot}(r/a)$, see Misra (1970).}
\begin{equation}  \label{V}
  V=-Mr/\Sigma.
\end{equation}
The structure of this field was shown in Semer\'ak (1995;
cf.\ Fig.\ 2 there) by depicting its lines in the
($\rho_{_{\rm f}},z$)-plane. We will use the Keres-Israel potential
as a scalar ``seed'' of our model. With $U=V$ given by Eq.\ (\ref{V}),
Eqs.\ (\ref{axisym-r})--(\ref{axisym-phi}) read
\begin{eqnarray}
  \Sigma\ddot{r}&=&
    m^{2}M\,(r^{2}+a^{2})(\Sigma-2r^{2})/\Sigma^{2}
     \nonumber \\ &~&
   -r(a\dot{r}\sin\theta)^{2}/(r^{2}+a^{2})
   +a^{2}\dot{r}\dot{\theta}\sin2\theta  \nonumber \\ &~&
   +r(r^{2}+a^{2})(\dot{\theta}^{2}+\dot{\phi}^{2}\sin^{2}\theta)
   \, , \label{V-r} \\
  \Sigma\ddot{\theta}&=&
   \{2m^{2}Ma^{2}r/\Sigma^{2}
   +a^{2}\dot{\theta}^{2}-a^{2}\dot{r}^{2}/(r^{2}+a^{2}) \nonumber \\ &~&
   +(r^{2}+a^{2})\dot{\phi}^{2}\}\sin\theta\cos\theta
   -2r\dot{r}\dot{\theta} \, , \label{V-theta} \\
  \ddot{\phi}&=&
   -2\,[r\dot{r}/(r^{2}+a^{2})+\dot{\theta}\cot\theta]\,\dot{\phi}
    \, . \label{V-phi}
\end{eqnarray}
Separated first integrals of these equations (analogy of Carter's
equations for the Kerr spacetime) were found by Israel
(1970). It was illustrated in Semer\'ak (1996) that Eqs.\
(\ref{V-r})--(\ref{V-phi}) often yield trajectories indiscernible from
their exact-Kerr counterparts computed from Eqs.\
(\ref{Kerr-r})--(\ref{Kerr-phi}). However, they do not contain the terms
linear in velocities which embody the very characteristic feature of the
Kerr geometry --- the frame-dragging effects.

\subsection{The dragging component}
\label{potentKerr-drag}
In this section, we are led by analogy with classical
electrodynamics and by observation that the Kerr dipole-like
gravitomagnetic field resembles the Kerr-Newman magnetic field.
One thus arrives at the potential which incorporates dragging in the form
\begin{equation}  \label{U1}
  U=V-\vec{v}\cdot\vec{A},
\end{equation}
where $\vec{v}=\vec{p}/m$ and the vector potential $\vec{A}$ is given by
that of the Kerr-Newman electromagnetic field,
$A_{i}=(0,0,Qra\sin^{2}\theta/\Sigma)$, with $Q$ (electric charge of the
Kerr-Newman centre) replaced by $-2M$ (the extra factor of 2 is ascribed
to the tensorial character of gravity and the minus sign to its
attractive nature) --- i.e.,
$A_{i}=(0,0,2Va\sin^{2}\theta)=(0,0,g_{t\phi})$. Similar (just half)
correction for the dragging was considered by Dadhich (1985) for a special
case of particles with zero axial angular momentum. The added term
$-\vec{v}\cdot \vec{A}$ really introduces the desired dragging
terms into the Eqs.\ (\ref{V-r})--(\ref{V-phi}): they read now
\begin{eqnarray}
  \Sigma\ddot{r}&=&
   [\mbox{r.h.s. of Eq.~}(\ref{V-r})] -2mM\Sigma^{-2}
   \nonumber \\ &~&
   \times (r^{2}+a^{2})(\Sigma-2r^{2})
   a\dot{\phi}\sin^{2}\theta \, , \label{U1-r} \\
  \Sigma\ddot{\theta}&=&
   [\mbox{r.h.s. of Eq.~}(\ref{V-theta})] \nonumber \\
   & & -2mMr\Sigma^{-2}(r^{2}+a^{2})a\dot{\phi}\sin2\theta
   \, , \label{U1-theta} \\
  {\half}(r^{2}+a^{2})\Sigma^{2}\ddot{\phi}&=&
   mMa\,[(\Sigma-2r^{2})\,\dot{r}  \nonumber \\ &~&
   +2r(r^{2}+a^{2})\,\dot{\theta}\cot\theta] \nonumber \\
   &~& -\Sigma^{2}
   [r\dot{r}+(r^{2}+a^{2})\,\dot{\theta}\cot\theta]\,\dot{\phi}
   \, . \label{U1-phi}
\end{eqnarray}
Note that the contravariant 3-potential obtained by using the
inversion of the Kerr 3-metric, $g^{ij}=1/g_{ij}$, is
\begin{equation}
A^{i}=(0,0,g_{t\phi}/g_{\phi\phi})=(0,0,-\omega_{_{\rm K}}),
\end{equation}
where $\omega_{_{\rm K}}=2Mar/{\cal A}$ and
${\cal{}A}=\Delta\Sigma+2Mr(r^{2}+a^{2})$; $A^{i}$ equals the shift
vector of Thorne et al. (1986) which stands for the
potential of the gravitomagnetic field in this reference.

\subsection{The space-geometry component}
\label{potentKerr-space}
The space curvature cannot be understood directly within the
Newtonian or electromagnetic analogy. It may only be included by
introducing ad hoc corrections into the form (\ref{U1}):
\begin{eqnarray}
 U & = & V-\vec{v}\cdot \vec{A}+\frac{(\vec{v}\cdot \vec{A})^{2}}{4V}
     -\frac{Mr\Sigma}{(r^{2}+a^{2})\Delta}\,
     \frac{\dot{r}^{2}}{m^{2}} \nonumber \\
   & = & -\frac{Mr}{\Sigma}\,
     \left(1-\frac{\dot{\phi}}{m}a\sin^{2}\theta\right)^{2}
    -\frac{Mr\Sigma}{(r^{2}+a^{2})\Delta}\,
     \frac{\dot{r}^{2}}{m^{2}} \, .   \label{U2}
\end{eqnarray}
This potential leads to equations of motion
\begin{eqnarray}
  \Sigma\ddot{r}&=&
   M\Delta\Sigma^{-2}(\Sigma-2r^{2})
   (m-a\dot{\phi}\sin^{2}\theta)^{2}
      \nonumber \\ &~&
   +[(r-M)\Sigma/\Delta-r]\,\dot{r}^{2}  \nonumber \\ &~&
   +a^{2}\dot{r}\dot{\theta}\sin2\theta
   +r\Delta(\dot{\theta}^{2}+\dot{\phi}^{2}\sin^{2}\theta)
   \, , \label{U2-r} \\
  \Sigma\ddot{\theta}&=&
   \{2Mr\Sigma^{-2}[am- \nonumber \\ &~&
   (r^{2}+a^{2})\dot{\phi}]^{2}
   +a^{2}\dot{\theta}^{2}
   \nonumber \\ &~&
   -a^{2}\dot{r}^{2}/\Delta
   +\Delta\dot{\phi}^{2}\}\sin\theta\cos\theta
   -2r\dot{r}\dot{\theta} \, , \label{U2-theta} \\
  {\half}\Sigma{\cal A}\ddot{\phi}&=&
   Ma(\Sigma-2r^{2})(m-a\dot{\phi}\sin^{2}\theta)\,\dot{r}
      \nonumber \\ &~&
   +2Mr(r^{2}+a^{2})[am-(r^{2}+a^{2})\dot{\phi}]\,
    \dot{\theta}\cot\theta  \nonumber \\ &~&
   -\Sigma^{2}(r\dot{r}+\Delta\dot{\theta}\cot\theta)\dot{\phi}
   \,. \label{U2-phi}
\end{eqnarray}

The only point in which Eqs.\ (\ref{U2-r}) and (\ref{U2-theta}) differ
from the relativistic Eqs.\ (\ref{Kerr-r})--(\ref{Kerr-theta}) is that
the relativistic variable $\dot{t}$ --- given by
$(\Delta\Sigma)^{-1}({\cal{A}}E-2Mar\Phi)$ ($E$ and $\Phi$ stand for the
particle's energy and axial angular momentum at infinity) --- is
replaced by $m$ in the classical model. This distinction reflects,
however, the conflict between the very roots of classical physics
(where time parameter is universal) and relativity (where proper time
and some coordinate time occur, related to each other in a specific way
at each point). Notice that Eq.\ (\ref{U2-phi}) differs from
(\ref{Kerr-phi}) also in several other points (namely $M$ is neglected 
three times).

\subsection{Specific features of motion in the pseudo-Kerr potential}
\label{motion}
Independence of the Lagrangian $L=T-mU$ on $t$ and $\phi$ implies two
constants of motion,
\begin{eqnarray}
 E &=& mv^{2}/2-mMr/\Sigma \nonumber \\
   &=& \frac{1}{2m}
   \left[\Sigma\left(\frac{\dot{r}^{2}}{\Delta}+\dot{\theta}^{2}\right)
   +\frac{{\cal{}A}}{\Sigma}\dot{\phi}^{2}\sin^{2}\theta\right]-
   \frac{mMr}{\Sigma}
    \,, \\
  \Phi &=& \frac{m{\cal{}A}}{\Sigma}\,\left(\omega-\omega_{_{\rm K}}\right)
    \sin^{2}\theta, \label{Phi}
\end{eqnarray}
where $\omega={\rm d}\phi/{\rm d}t=\dot{\phi}/m$. The Kerr axial
angular momentum at infinity contains (again, as in Eqs.\
(\ref{U2-r})--(\ref{U2-theta})) $\dot{t}$ instead of $m$ ($\omega_{_{\rm
K}}$ is the angular velocity with which the field rotates relative to an
observer standing at infinity). For large $r$ one obtains the usual forms of
the energy and axial angular momentum in the Newtonian central
gravitational field:
\begin{eqnarray}
  E&\simeq &(2m)^{-1}(\dot{r}^{2}+r^{2}\dot{\theta}^{2}
             +r^{2}\dot{\phi}^{2}\sin^{2}\theta)-mM/r, \\
  \Phi&\simeq &mr^{2}\dot{\phi}\sin^{2}\theta.
\end{eqnarray}

Let us determine the acceleration $a^{i}$ of an observer orbiting
uniformly ($\omega={\rm const}$) at $r={\rm const}$, $\theta={\rm
const}$. We will define acceleration as a specific force necessary to
keep the observer in the orbit, i.e.\ as the {\em minus} acceleration of
a {\em free} particle having $\dot{r}=\dot{\theta}=0$ at a given
point. According to Eqs.\ (\ref{U2-r})--(\ref{U2-phi}), we find
\begin{eqnarray}
  a^{r}&=&\Delta\Sigma^{-3}
    [M(2r^{2}-\Sigma)(1-a\omega\sin^{2}\theta)^{2}
       \nonumber \\ &~&
    -    r(\Sigma\omega\sin\theta)^{2}] \, , \label{ar} \\
  a^{\theta}&=&-\Sigma^{-3}
    \{2Mr[a-(r^{2}+a^{2})\omega]^{2}
       \nonumber \\ &~&
    +\Delta\Sigma^{2}\omega^{2}\}
    \sin\theta\cos\theta \, , \label{atheta} \\
  a^{\phi}&=&0 \label{aphi} \, .
\end{eqnarray}
This 3-acceleration differs from the space part of 4-acceleration of the
Kerr stationary observer only by the absence of the multiplicative
factor $(u^{t})^{2}$ (square of the time-component of the observer's
4-velocity) [cf.\ Semer\'ak 1993, Eqs. (37)--(39)].

For a static observer ($\omega=0$) one obtains, in particular,
\begin{equation}  \label{aSO}
  a^{i}=(M/\Sigma^{3})
        \left(\Delta(2r^{2}-\Sigma),-ra^{2}\sin2\theta,0\right).
\end{equation}
Hence, the field is repulsive at $0\leq r<a|\cos\theta|$ in the
sense that the radial component of (\ref{aSO}) is negative there.

According to (\ref{ar})--(\ref{aphi}), the stationary observer needs no
thrust (i)~at $r=a$ on the axis ($\theta=0^{\circ},180^{\circ}$), and
(ii)~in the equatorial plane if his angular velocity is
$\omega=1/(a\pm\sqrt{r^3/M})$. The latter agrees exactly
with the Keplerian angular velocity of an equatorial observer orbiting
along a circular geodesic in the Kerr spacetime.

Important circular geodesics in the equatorial plane, on the other hand,
are not reproduced properly by the potential (\ref{U2}), viz. the equation for marginally stable orbits reads
\begin{equation}
  r^{3}\pm 8a\sqrt{Mr^{3}}-3a^{2}r-2Ma^{2}=0
\end{equation}
and for marginally bound orbits
\begin{equation}
  r^{2}\pm 4a\sqrt{Mr}-a^{2}=0.
\end{equation}
The correct equations have respectively the forms
\begin{equation}
  r^{2}-6Mr\pm 8a\sqrt{Mr}-3a^{2}=0
\end{equation}
and
\begin{equation}
  r^{2}-4Mr\pm 4a\sqrt{Mr}-a^{2}=0.
\end{equation}
In the Schwarzschild case, for instance, both our equations imply $r=0$.

In any pseudo-Newtonian description of a rotating black hole, the main
difficulty is to simulate the presence of the horizon and of the
dragging effects with an acceptable precision. Both these phenomena are
outside the scope of Newtonian physics. Our potential accounts for the
frame-dragging effects, and especially in the intermediate and large
distances provides a good fit, whereas it does not describe correctly
the innermost region where the horizon and important circular orbits
lie. In order to account for the horizon, one would have to start from
the scalar potential which diverges to $-\infty$ there. The Keres-Israel
potential (\ref{V}), instead, reaches $-\infty$ only at the very
singularity $\Sigma=0$. The potential proposed by Artemova et al.\
(1996) is a better alternative in this respect, being the simplest
generalization of the Paczy\'nski-Wiita potential to the Kerr case which
reproduces the horizon and approximates the important orbits. Also, the
epicyclic frequency of small radial oscillation $\kappa(r)$, important in
the theory of discoseismology, is not reproduced with good accuracy by
potentials (\ref{V}) and (\ref{U2}) although it does show a maximum,
typical for a relativistic $\kappa$. Apparently, trying to incorporate
different manifestations of the frame-dragging accurately, one may end
with a long expression without practical sense.

To summarize, each particular relativistic effect can
be well simulated within classical physics, but pseudo-Newtonian
modelling of the complete relativistic situation has only a restricted
validity. In order to clarify the value of our potential, we have
carried out an extensive computation of trajectories and introduced a
criterion which determines the accuracy of the pseudo-Newtonian model.

\section{Testing approximative equations}
\label{test}
As mentioned above, the pseudo-Newtonian potential has been used
frequently in various situations in which general relativistic effects
on the motion of material (either test particles or fluids) are
essential but exact calculations are too difficult. It is however
impossible to estimate, {\it a priori}, the error which is introduced by
replacing the original system, described in the framework of general
relativity, by a corresponding pseudo-Newtonian system. The plausibility of
a particular form of the simulating potential can be verified by solving
analogous situations within the exact theory. For example, several specific
questions in the astrophysics of accretion discs had been first analysed
using the pseudo-Newtonian theory, and the results were only later
supported by more complicated calculations within the Schwarzschild
spacetime (with identical local physics of the fluid). Indeed, it is
quite trivial to recall that the above-mentioned standard model of thin
discs was formulated within the Newtonian (Shakura \& Sunyaev 1973),
pseudo-Newtonian (Paczy\'nski \& Wiita 1980), and also relativistic
frameworks (Novikov \& Thorne 1973). A specific response of relativistic
accretion discs to oscillations (Kato \& Fukue 1980) was also explored
with a modified Newtonian potential (Nowak \& Wagoner 1991) before
further steps were carried out (Perez et al.\ 1997). These examples
indicate that the pseudo-Newtonian model is appropriate for
investigating the motion of material around black holes.

It is quite straightforward to check whether the pseudo-Newtonian
potential is suitable for treating the motion of test particles and
fluids when one deals with a spherically symmetric system. What appears
more difficult is to develop an analogous pseudo-Kerr theory, describing
the relativistic effects near a rotating compact object. Now we will
propose a systematic approach to estimate the quality of a particular
model by computing a sufficiently large number of trajectories with
different initial conditions and comparing results with
exact calculations of geodesic motion. The test we suggest determines
the rate of divergence of trajectories (as given by approximative
versus exact equations). We propose that this type of check
should be carried out for each particular set of approximative equations
before they are applied to astrophysical situations. (Until now, only
rather restricted tests on a relatively small number of trajectories have
been applied. As a consequence, one cannot be sure about predictions
based on approximative equations of such models because there is little
or no control of their overall precision.) For illustration, we then submit the
approximative equations of Sect.\ \ref{potentKerr} to our test.

\subsection{Criterion of accuracy of approximative solutions}
Our criterion is motivated by the definition of the Lyapunov characteristic
coefficients which have been introduced in order to characterize chaotic
systems (see, e.g., Chapt.~5 of Lichtenberg \& Lieberman (1983).
The most important features of the Lyapunov coefficients relevant for
the present work are summarized in the Appendix.

We will now introduce a parameter which is analogous to the maximum
Lyapunov characteristic exponent $\lambda$. One should however realize
the basic difference between the standard formulation of the problem of
chaotic system and our present situation. While $\lambda$
characterizes the rate of divergence of two neighbouring (initially
close to each other) trajectories, in the present work one always starts
with {\it exactly identical\/} initial conditions. Trajectories then
separate because the motion of the first test particle is determined by
the geodesic equation in the Kerr spacetime, while the other
(fictitious) particle moves according to approximative equations.
The geodesic equation could be integrated
analytically but not the approximative equations, so we have to
resort to numerical solution. Two points should be mentioned: (i)~as the
test particles move in a stationary system with preferred time
coordinate ($t$), the separation of trajectories in the phase space is
calculated in the $t=\const$ slice; (ii)~periodic rescalings of the
separation $w$ to zero length keep the two trajectories close to each
other during their evolution.

In analogy with standard definition (\ref{lambda}), we introduce a
critical parameter
\beq
\Lambda=
\lim_{n\rightarrow\infty}(n\Deltat)^{-1}\sum_{k=1}^{n}\ln
\frac{d_k}{d_0},
\label{Lambda}
\eeq
with $d_0\equiv|w_0(\Deltat)|.$

Now:
\begin{description}
\item[(i)]~two particles start with identical initial conditions;
the first one evolves according to geodesic equations
(\ref{Kerr-r})--(\ref{Kerr-phi}), while the second one according to
approximative equations (Eqs.\ (\ref{U2-r})--(\ref{U2-phi}), for example);
\item[(ii)]~$\Deltat$ is determined as a
time interval during which the two trajectories remain close to each
other and the initial separation $d_0$ is an arbitrary small number
(typically, $\Deltat$ is of
the order of the Keplerian orbital period at the initial radius);
\item[(iii)]~Equation~(\ref{Lambda}) is evaluated and it is determined,
numerically, whether convergence has been reached for large $n$ with a
pre-determined accuracy. The final value, $\Lambdaf$, characterizes
the rate of divergence of the two trajectories with the same initial
conditions: $\Lambdaf>0$ corresponds to separation increasing
exponentially while $\Lambdaf\rightarrow-\infty$ corresponds to a
polynomial (i.e.\ much slower) increase of the separation.
\end{description}

In order to estimate the accuracy of the approximative equations, one needs
to follow a large number of trajectories with randomly chosen initial
conditions. By averaging over $\Lambdaf$ corresponding to different
initial conditions, one obtains a value which shows whether the
approximative equations describe, {\em on the whole}, the motion of
test matter with good precision. Regions where the mean value is
positive, $\langle\Lambdaf\rangle>0$, indicate that the approximation is
not acceptable. The process is described in more detail below where we
calculate $\langle\Lambdaf\rangle$ for Eqs.\
(\ref{V-r})--(\ref{V-phi}) and (\ref{U2-r})--(\ref{U2-phi}) as
examples.

\subsection{An example: Test of the pseudo-Kerr equations}
We submitted our approximative equations to the above-described test on
$\langle\Lambdaf\rangle$. Relative position of the two test particles is
determined by difference equations:
\beq
 \delta\dot{x}^i\equiv\dot{x}^i_{\rm{}Kerr}-\dot{x}^i_{\rm{}approx},\quad
 \delta\ddot{x}^i\equiv\ddot{x}^i_{\rm{}Kerr}-\ddot{x}^i_{\rm{}approx},
 \label{deltax}
\eeq
with $x^i=\{r(t),\theta(t),\phi(t)\}.$ The suffix ``Kerr''
indicates that the trajectory is a geodesic in the Kerr spacetime while
the suffix ``approx'' corresponds to pseudo-Newtonian equations.
We should stress in this place that comparing processes in
different spacetimes {\em is} a serious problem in general
relativity. Here, however, we do not compare two relativistic
spacetimes and the situation is quite different: though a proper
physical justification gives an additional appeal to any
pseudo-Newtonian model and we have therefore emphasized also a physical
content of our approach (in Sect.\ \ref{potentKerr}), in the final stage one
mainly asks whether the test particles in the model field in
{\em some} coordinates move along trajectories that are
sufficiently close to {\em certain} geodesics of a given relativistic
field in some particular coordinates. It is only necessary to
define the way of correspondence of the initial conditions. In
our test, we consider as counterparts the particles which start
from a given position ($r$,$\theta$) with given (specific) constants of
motion ($e=E/m$,$l=\Phi/m$). Of course, it is possible that we would
obtain a better fit with some other choice, e.g. if the
``corresponding'' particles had the same initial velocities with
respect to some (corresponding) local frames.

\begin{figure}
\epsfxsize=1.02\hsize
\centering
\mbox{\epsfbox{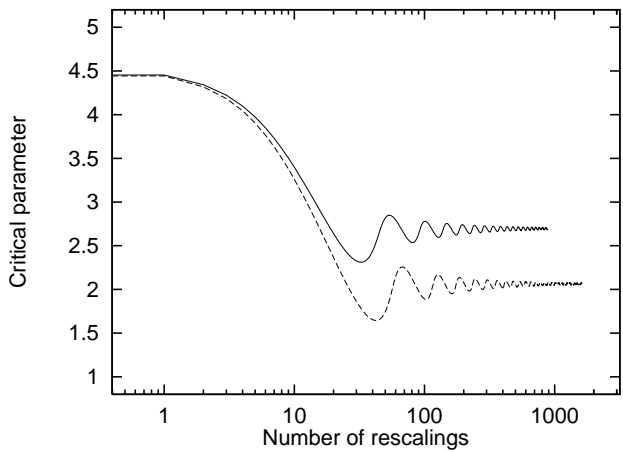}}
\caption{This graph illustrates how the critical parameter $\Lambda$
oscillates at first and then settles to a positive value
after $k\approx10^3$ rescalings, indicating the chaotic-type increase of
separation $d_k$ (cf.\ Eq.~(\protect\ref{Lambda})). Two different
examples of trajectories are shown.
\label{fig1}}
\end{figure}

\begin{figure}
\epsfxsize=1.07\hsize
\centering
\mbox{\epsfbox{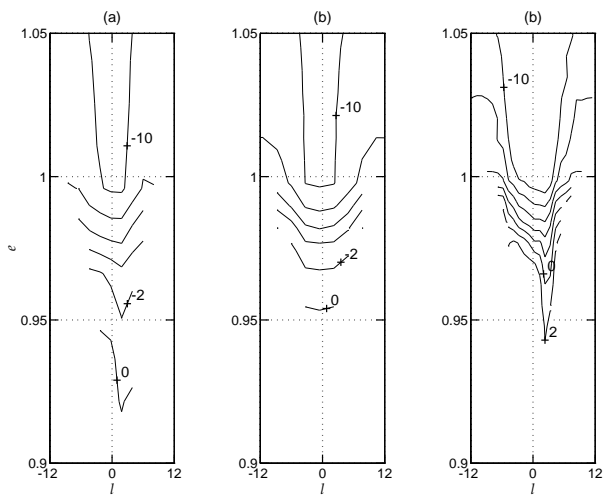}}
\caption{Isocurves of the mean terminal value of the critical
parameter, $\langle\Lambdaf\rangle$, as a function of the particles'
specific energy $e$ and their specific axial
angular-momentum $l$. Three panels correspond to (a)~Eqs.
(\protect\ref{U2-r})--(\protect\ref{U2-phi}) with $a=M$, (b)~Eqs.
(\protect\ref{U2-r})--(\protect\ref{U2-phi}) with $a=0$, and, for
comparison, (c)~Eqs. (\protect\ref{V-r})--(\protect\ref{V-phi}) with $a=M$
(scalar, Keres-Israel model). Positive $\langle\Lambdaf\rangle$ indicates
chaotic-type behaviour (see the text for details).
\label{fig2}}
\end{figure}

We have integrated difference Eqs. (\ref{deltax}) using the
Bulirsch-Stoer scheme (Press et al.\ 1994). Initial conditions
cover the parameter space of
\beq
 r>r_+\equiv M+\sqrt{M^2-a^2},
 \; 0<\theta\lta\half\pi,\; e<e_{\rm max}.
\label{ic}
\eeq
Typically, we set $e_{\rm max}\approx1.1$. As expected,
most of the particles with $e_{\rm max}\gg1$ escape quickly to large
distances $(r\gg r_+)$ where both the exact and approximative equations
give identical results. It is thus relevant to investigate trajectories
with lower energies which often make a number of revolutions around the
black hole. Some of these trajectories tend to diverge in a chaotic-type
manner, as we will see below.

Each run, characterized by the value of $a/M$, resulted in about $10^7$
values of $\Lambdaf$. For each set of initial conditions (\ref{ic})
within the run, the separation of the two corresponding trajectories was
periodically rescaled down to a pre-determined value after a fixed
interval $\Deltat$. We have typically chosen $\Deltat=2M$ which is
comparable to the Keplerian orbital period near the horizon. The integration
was terminated when one of the following conditions had been satisfied:
(i)~$\Lambda$ converged to a finite positive value $\Lambdaf$
(numerically, we checked that the relative change of $\Lambda$ during
the last 30 rescalings did not exceed 0.5\,\%); (ii)~the upper limit of
$10^4$ on the number of rescalings was exceeded, while $\Lambda$ was
oscillating or decreasing monotonically to negative values; (iii)~the
trajectory was captured by a black hole, $r\leq{}r_+.$ The case (iii)
was excluded from further consideration.

Figure~\ref{fig1} illustrates the evolution of $\Lambda$ for two sets of
initial conditions which both fall under item~(i) above. Apparently,
both cases shown there converge to positive $\Lambdaf$, although the
numbers of rescalings were different. In this respect, it is interesting
to note that the separation of nearby geodesics in the Kerr spacetime
never increases exponentially. This fact is a consequence of the
existence of the fourth (Carter's) constant of motion, as discussed in
Karas \& Vokrouhlick\'y (1992).

Next we illustrate the mean value $\langle\Lambdaf\rangle$ as a
function of the constants of motion, $e$ and $l$. Graphs corresponding
to Eqs.\ (\ref{U2-r})--(\ref{U2-phi}) are shown for an extremely rotating
black hole (Fig.~\ref{fig2}a) and a non-rotating black hole
(Fig.~\ref{fig2}b). For comparison, the Keres-Israel model
(\ref{V-r})--(\ref{V-phi}) with $a=M$ was also included
(Fig.~\ref{fig2}c). One can locate easily the regions of
positive $\langle\Lambdaf\rangle$. In general these regions are bound to
small $e$ and $l$ which correspond to trajectories that plunge close to
the horizon. In the case of a non-rotating hole the graph is symmetrical
about $l=0$ because both the Schwarzschild spacetime and the
adopted pseudo-Newtonian field with $a=0$ are spherically
symmetric. Distortion introduced by rotation is visible in the graphs
(a) and (c). Comparing these two graphs one can verify that model
(\ref{U2}) is superior to (\ref{V}) (cf.\ the positive values of
$\Lambdaf$ in Fig.\ \ref{fig2}c, indicating that most of the
corresponding trajectories separate rapidly from each other).

\section{Conclusions}
We have described a method to test the overall accuracy of the
pseudo-Kerr models of the relativistic gravitational field and we
applied our approach to two particular potentials, given by Eqs.\
(\ref{V}) and (\ref{U2}). One can conclude that the corresponding
equations of motion, (\ref{V-r})--(\ref{V-phi}) and
(\ref{U2-r})--(\ref{U2-phi}) respectively, do not provide an acceptable
approximation to exact Kerr geodesic equations in the part of the
$(e,l)$-plane where $\langle\Lambdaf\rangle>0$, and vice-versa. This
statement does not apply strictly to {\em all} trajectories (with
different initial conditions) because our graphs are given in terms of
mean values. We suggest that a check of this type should always
be employed when some particular form of approximative equations is
formulated. For example, calculation of the spectra of accretion discs
requires integration of a large number of photon trajectories; the
overall accuracy in the description of the gravitational field is then
important.

\begin{acknowledgements}
We thank for support from the grants
GACR-202/\linebreak[2]96/\linebreak[2]0206 of the Grant Agency of the
Czech Republic and GAUK-36/97 from the Grant Agency of the Charles
University.

\end{acknowledgements}

\appendix
\section{Appendix}
The Lyapunov characteristic exponent $\lambda$ of two nearby
trajectories, $\ell$ and $\ell+w$, is defined by
\beq
\lambda(\ell,w)=
\lim_{\stackrel{t\rightarrow\infty}{d(0)\rightarrow0}}
 t^{-1}\ln\left[\frac{d(t)}{d(0)}\right],
\eeq
where $\ell(t)$ is a solution of the equations of motion in the phase
space,
$w(t)$ is a connecting vector, $d(t)=|w|$ is its length in the phase
space and $d(0)$ is an initial separation (the value of $d(0)$ must be
small; otherwise it is arbitrary and one usually scales $d(0)$ to unity).
The Kerr spacetime being stationary, we can measure the separations
on the $t={\rm const}$ surfaces. One defines $\lambda_{\rm max}$ as the
maximum value of $\lambda$ with respect to variations of $w$ and
characterizes the chaoticity of the system by the value of
$\lambda_{\rm{}max}$.
Positive values of $\lambda_{\rm max}$ indicate that neighbouring
trajectories diverge exponentially in the course of time while negative
$\lambda_{\rm max}$ corresponds only to polynomial divergence. The above
definitions have been originally introduced within the framework of
non-relativistic systems but they are directly applicable also to
stationary systems in general relativity.

The maximum Lyapunov characteristic exponent is frequently determined
numerically. This approach requires a careful choice of the
integration scheme. In order to keep computational errors under control,
one lets the trajectories evolve for a short interval of time,
$\Deltat$, after which $w$ is rescaled back to unity (one denotes
$d_k\equiv|w_{k-1}(\Deltat)|$ the norm of the connecting vector at the
moment of the $k$-th rescaling). One can show (Benettin 1984) that the
Lyapunov characteristic exponent corresponding to the original $w_0$ is
given by
\beq
\lambda=
\lim_{n\rightarrow\infty}(n\Deltat)^{-1}\sum_{k=1}^{n}\ln
\frac{d_k}{d_0},
\label{lambda}
\eeq
independently of the value of $\Deltat.$ In addition,
$\lambda=\lambda_{\rm max}$ for almost all $w_0.$ Again, one can
conveniently set $d_0=1$.


\begin{thebibliography}{99}
\bibitem{ACLS88}
   Abramowicz M.A., Czerny B., Lasota J.P., Szuszkiewicz E.,
   1988, ApJ 332, 646
\bibitem{ABChI96}
   Abramowicz M. A., Beloborodov A.M., Chen X.-M.,
   Igumenshchev, I.V., 1996, A\&A 313, 334
\bibitem{ABN96}
   Artemova I.V., Bj\"ornsson G., Novikov I.D., 1996,
   ApJ 461, 565
\bibitem{B70}
   Bardeen J. M., 1970, Nat 226, 64
\bibitem{BP75}
   Bardeen J.M., Petterson, J.A., 1975, ApJ 195, 165
\bibitem{B76}
   Benettin G., 1984, Physica D 13, 211
\bibitem{ChK92}
   Chakrabarti S.K., Khanna R., 1992, MNRAS 256, 300
\bibitem{D85}
   Dadhich N., 1985,
   in: N. Dadhich, C.V. Vishveshwara (eds.),
   {\it A Random Walk in Relativity and Cosmology}.
   Wiley Eastern, New Delhi, p. 72
\bibitem{DM97}
   Daigne F., Mochkovitch R., 1997, MNRAS 285, L15
\bibitem{I70}
   Israel W., 1970, Phys. Rev. D 2, 641
\bibitem{IFRN96}
   Iwasawa K., Fabian A.C., Reynolds C.S., et al.,
   1996, MNRAS 282, 1038
\bibitem{J67}
   Jaen J. K., 1967, Phys. Rev. 163, 1361
\bibitem{JCB92}
   Jantzen R. T., Carini P., Bini D., 1992,
   Ann. Phys. (N.Y.) 215, 1
\bibitem{KK96}
   Karas V., Kraus P., 1996, PASJ 48, 771
\bibitem{KV92}
   Karas V., Vokrouhlick\'y D., 1992, Gen. Rel. Grav. 7, 729
\bibitem{KF80}
   Kato S., Fukue J., 1980, PASJ 32, 377
\bibitem{KFM98}
   Kato S., Fukue J., Mineshige S., 1998,
   {\it Black-Hole Accretion Disks}. Kyoto University Press, Kyoto
\bibitem{K67}
   Keres H., 1967, Zh. eksp. teor. fiz. 52, 768
\bibitem{Kr80}
   Krasi\'nski A., 1980, Phys. Lett. 80A, 238
\bibitem{LL83}
   Lichtenberg A.J., Lieberman, M.A., 1983,
   {\it Regular and Stochastic Motion}. Springer-Verlag, New York
\bibitem{ML98}
   Markovi\'c D., Lamb F.K., 1998, ApJ, in press;
   also astro-ph/9801075
\bibitem{Mi70}
   Misra M., 1970, Proc. Roy. Ir. Acad. 69A, 39
\bibitem{MFWK98}
   Miwa T., Fukue J., Watanabe Y., Katayama M., 1998, PASJ,
   in press
\bibitem{NT73}
   Novikov I.D., Thorne K.S., 1973, in:
   DeWitt C., DeWitt B.S. (eds.),
   {\it Black Holes}. Gordon and Breach, New York, p. 343
\bibitem{NW91}
   Nowak M.A., Wagoner R.V., 1991, ApJ 378, 656
\bibitem{OKF87}
   Okazaki A.T., Kato S., Fukue J., 1987, PASJ 39, 457
\bibitem{PW80}
   Paczy\'nski B., Wiita P.J., 1980, A\&A 88, 23
\bibitem{PSWL97}
   Perez C.A., Silbergleit A.S., Wagoner R.V., Lehr D.E.,
   1997, ApJ 476, 589
\bibitem{PTVF94}
   Press W.H., Teukolsky S.A., Vetterling W.T.,
   Flannery B.P., 1994, {\it Numerical Recipes}.
   Cambridge Univ. Press, Cambridge
\bibitem{Q86}
   Qadir A., 1986, Europhys. Lett. 2, 427
\bibitem{R98}
   Rees M.J., 1998, in: R.M. Wald (ed.),
   {\it Black Holes and Relativistic Stars}.
   Univ. Chicago Press, Chicago
\bibitem{RJ98}
   Ruffert M., Janka H.-Th., 1998, A\&A, submitted;
   also astro-ph/9809280
\bibitem{S93}
   Semer\'ak O., 1993, Gen. Rel. Grav. 25, 1041
\bibitem{S95}
   Semer\'ak O., 1995, Nuovo Cim. B 110, 973
\bibitem{S96}
   Semer\'ak O., 1996, Astrophys. Lett. Commun. 33, 275
\bibitem{SS73}
   Shakura N. I., Sunyaev R.A., 1973, A\&A 24, 337
\bibitem{SV98}
   Stella L., Vietri M., 1998, ApJ 492, L59
\bibitem{SM98}
   Szuszkiewicz E., Miller J.C., 1998, MNRAS 298, 888
\bibitem{TPM86}
   Thorne K.S., Price R.H., Macdonald D.A., eds., 1986,
   {\it Black Holes: The Membrane Paradigm}.
   Yale Univ. Press, New Haven, Chapt. III.\,A
\bibitem{W98}
   Wagoner R.V., 1998, Phys. Rep., in press;
   also astro-ph/9805028
\bibitem{WZCh98}
   Wei Cui, Zhang S.N., Chen W., 1998, ApJ 492, L53

\end{thebibliography}
\end{document}